\documentclass[12pt]{JHEP3}
\usepackage{graphicx,amsmath,times,cite}
\newcommand{\link}[1]{\langle#1\rangle}
\newcommand{\tr}{\text{tr}}
\newcommand{\Disc}{\text{Disc}}
\newcommand{\bL}{\mathfrak{b}}

\newcommand{\FIG}[3]{
\FIGURE[t]{\includegraphics[width=14cm]{#1}\caption{#2}\label{#3}}}

\preprint{IPhT-T08/193 \\ KIAS-P08079}
\title{Boundary operators in the O(n) and RSOS matrix models}
\author{Jean-Emile Bourgine$^1$ and Kazuo Hosomichi$^2$ \\ \\
$^1$ Institut de Physique Th\'eorique, CNRS-URA 2306  \\ ~~
     C.E.A.-Saclay, F-91191 Gif-sur-Yvette, France    \\
$^2$ Korea Institute for Advanced Study, Seoul 103-722, Korea \\
\email{jean-emile.bourgine@cea.fr},
\email{hosomiti@kias.re.kr}
}

\abstract{
We study the new boundary condition of the $O(n)$ model proposed
by Jacobsen and Saleur using the matrix model.
The spectrum of boundary operators and their conformal weights
are obtained by solving the loop equations.
Using the diagrammatic expansion of the matrix model as well as
the loop equations, we make an explicit correspondence
between the new boundary condition of the $O(n)$ model
and the ``alternating height'' boundary conditions in RSOS model.
}
\keywords{Matrix theory, Noncritical string theory}

\begin{document}

\section{Introduction}

Boundary conformal field theories play an important role
in many fields of theoretical physics, such as
statistical mechanics, condensed matter or string theory.
In order to study the properties of boundary conditions
and boundary operators, it is useful to have at our disposal
a microscopic description of the conformal field theories (CFT).
The $O(n)$ model and solid-on-solid (SOS) models are
the familiar examples which provide us with such a description
of, in general irrational, CFTs with central charge $c<1$.
In the $O(n)$ model each lattice site is assigned an $O(n)$ spin,
whereas in SOS models one associates an integer-valued height
to each lattice point.
In both theories, neighbouring sites are then coupled
via suitable interactions.
The heights are bounded from both sides in the so called
restricted SOS (or RSOS) models; these models are known to describe
rational CFTs with $c<1$.

Both the $O(n)$ and SOS models can be
reformulated as loop gas models \cite{Kostov:1991cg}.
In this formulation, the $O(n)$ model makes sense for
arbitrary real $n$ and exhibit critical behaviour for $|n|\le2$.
The phase structure of these models is well understood.
Interestingly, they are known to describe two CFTs of different
central charges connected by a renormalization group flow.
The loops behave differently in the UV (or {\it dilute})
phase and the IR ({\it dense}) phase.

Some properties of the $O(n)$ and SOS models can be studied
by putting them on a fluctuating lattice, i.e.,
coupling them to the two-dimensional gravity.
The partition function of such theories is given by
summing up the partition functions of the model
on all the different lattices weighted by their area.
Actually, the $O(n)$ and SOS models on random
-- or dynamical -- lattice are known to be described by
the Feynman graph expansion of certain matrix models
\cite{Kostov:1991cg,Kazakov:1986hy,Boulatov:1986sb,Duplantier:1988wc,
      Kazakov:1988fv,Kostov:1988fy,Kostov:1989eg}.
In this context, the continuum limit is achieved by tuning the
potential couplings while sending the size of the matrices to infinity.
In this limit one recovers the dynamics of the irrational CFT with
$c<1$ coupled to the Liouville gravity and reparametrization ghosts.

The $O(n)$ and SOS matrix models are also useful
in studying the conformally invariant boundary conditions
from the microscopic viewpoint.
In this paper, we will be particularly interested
in the boundary conditions of the $O(n)$ model
recently proposed by Jacobsen and Saleur \cite{Jacobsen:2006bn}.
Instead of allowing the loops to touch the boundary freely,
they weighted the loops touching the boundary differently
from those which do not.
They also considered the ``$L$-leg''
boundary operators on which $L$ open lines end.
The properties of such boundary conditions and operators were studied
on a fixed annular lattice with $L$ non-contractible loops introduced.
They obtained a continuous spectrum of boundary operators
and determined their conformal weights.
The new boundary conditions were also put on a dynamical lattice
by Kostov \cite{Kostov:2007jj}, where the correlation functions
and the conformal weights of the $L$-leg operators were computed.
The $L$-leg operators were also considered in some earlier
works \cite{Kazakov:1991pt,Kostov:2003uh,Kostov:2002uq}.

Another interesting fact is that the $O(n)$ model becomes
equivalent to the RSOS model for some special values of $n$.
In \cite{Jacobsen:2006bn} it was proposed that the new boundary conditions
of Jacobsen and Saleur correspond to the boundary conditions
in RSOS model which force the boundary height to alternate
between two values \cite{Saleur:1988zx}.

In this paper we study the property of boundary conditions
and boundary operators of these models using the loop equations
along the line of \cite{Kostov:2007jj}, but with more help of
the matrix model formulation which is much simpler to handle
than the combinatorics employed in the earlier work.
We will be focusing on the dense phase, leaving
the analysis of the dilute phase as a future work.

The organization of this paper is as follows.
In section \ref{sec:On} we introduce the new boundary conditions
and boundary $L$-leg operators in the $O(n)$ matrix model following
\cite{Kostov:2007jj}, and rederive the correlation functions and
conformal weights of the $L$-leg operators from the loop equations.
Then in section \ref{sec:sos} we propose a description of
the boundary conditions of alternating heights in RSOS matrix models.
Using this we derive the spectrum of boundary operators
as well as their conformal weights and correlators, again by
solving the loop equations.
Finally, in section \ref{sec:map} we show the equivalence
of the $O(n)$ and RSOS models on discs
by establishing a map between their Feynman graphs.
We use this to derive some relations between the disc
correlators, which are then shown to map the loop equations of
one theory to the other.
The last section \ref{sec:concl} is devoted to some conluding remarks.

In Appendix \ref{sec:kpz} we review some basic facts on the
Liouville theory approach to conformal field theories coupled
to two-dimensional gravity, and summarize the formulae
for the conformal weight and gravitational dimension of
the operators.
Some detail of solving the loop equation and reading off the
gravitational dimension are given in Appendix \ref{sec:sol}.

\section{Boundary operators in the $O(n)$ model}\label{sec:On}

\subsection{Definition of the model}\label{sec:defOn}

Let us consider a triangular lattice $\Gamma$ with
an $O(n)$ spin component associated to each site $r$,
normalized so that $\tr\,S_a(r)S_b(r')=\delta_{ab}\delta_{rr'}$.
The partition function of the $O(n)$ model is defined by
\cite{Nienhuis:1982fx, Nienhuis:1984wm}
\begin{equation}
 Z_\Gamma(T)=\tr\prod_{\link{rr'}}
 \left(1+\frac1T\sum_a S_a(r)S_a(r')\right),
\end{equation}
where $T$ is called the temperature and the product
runs over all links $\link{rr'}$ of $\Gamma$.
Expanding the product into a sum of monomials,
the partition function can be written as a sum over all
configurations of self avoiding, mutually avoiding loops on $\Gamma$,
\begin{equation}
Z_\Gamma(T)=\sum_{\text{loops}}T^{-(\text{length})}n^{\#(\text{loops})}.
\end{equation}
Each loop is counted with a factor $n$, and the temperature
$T$ controls the average total length of the loops.
When formulated in this way, the model
makes sense for arbitrary real $n$.
This model is known to exhibit a critical behaviour for $|n|\le2$.
Hereafter we parameterize $n$ in terms of $g$ or $\theta$ as follows,
\begin{equation}
n=-2\cos{(\pi g)}=2\cos{(\pi\theta)}.
\end{equation}
As a function of $n$, $g$ is multi-valued.
Different branches are known to correspond to
different phases of the model \cite{Kostov:1991cg}.

The temperature controls the phase of the model.
It is in the dilute phase at some critical temperature $T=T^*$,
and below $T^*$ it is in the dense phase.
For generic $n$, the two phases are described by two irrational
conformal field theories with central charges
\begin{equation}
c_\text{dense}=1-\frac{6\theta^2}{1-\theta},\qquad
c_\text{dilute}=1-\frac{6\theta^2}{1+\theta}.
\end{equation}
We will be focusing on the physics in the dense phase,
which has the same behaviour as the fully packed loop model
corresponding to $T=0$.
Hereafter we assume $\theta\in [0,1]$ and $\theta=1-g$.

Turning on the gravity corresponds to taking the sum
over all the triangulated surfaces with a suitable weight,
\begin{equation}
 Z_{\text{dyn}}(\kappa,T)=\sum_\Gamma\kappa^{-A(\Gamma)}Z_\Gamma(T).
\end{equation}
The parameter $\kappa$ controls the average
area $A(\Gamma)$ (the number of triangles)
and is regarded as the bare cosmological constant.
The continuum limit is obtained from the vicinity
of the critical line $\kappa=\kappa^*(T)$
where the average area of the surface diverges.
One can also allow the surfaces to have boundaries.
For example, a disc partition function can be defined
as the sum over the surfaces of disc topology,
\begin{equation}
 Z_{\text{dyn}}(\kappa,x,T)=
 \sum_{\Gamma:\;\text{disc}}\frac1{L(\Gamma)}
 \kappa^{-A(\Gamma)}x^{-L(\Gamma)}Z_\Gamma(T),
\end{equation}
where $x$ is the boundary cosmological constant controlling
the average boundary length (the number of edges along the boundary).
In this case, we have to send $x$ also to a critical value as
$\kappa\to\kappa^*(T)$ so that the average boundary length
diverges in the limit.
The continuum limit is therefore parametrized by
the renormalized couplings $\mu\sim\kappa^*-\kappa$
and $\xi\sim x-x^*$.
Note that, when the disc has more than one boundary,
a boundary cosmological constant may be introduced for each.

Until recently, the only boundary condition
studied in the $O(n)$ model was the Neumann boundary condition
in which the spins at the boundary fluctuate freely.
Based on earlier work \cite{Nichols:2004fb,Pearce:2006sz},
Jacobsen and Saleur \cite{Jacobsen:2006bn}
proposed a new kind of boundary conditions in which the boundary
spins are forced to take the first $k$ of the $n$ values.
We call this the $k$-th JS boundary condition.
Neumann and Dirichlet boundary conditions correspond
to the special cases with $k=n$ and $k=1$, respectively.
In the loop gas picture, the $k$-th JS boundary condition amounts
to giving a weight $k$, instead to the usual $n$, to the
loops that touch the boundary at lease once.
Defined in this way, the JS boundaries make sense for non-integer $k$.

Following \cite{Kostov:2007jj}, we consider the model on the disc
with one Neumann and one JS boundaries connected by the boundary
changing operators,
\begin{equation}
 \mathbb{S}_L^\parallel = 
 \sum_{1\leq a_1<\cdots<a_L\leq k}S_{a_1}\cdots S_{a_L},
 \qquad
 \mathbb{S}_L^\perp =
 \sum_{k<a_1<\cdots<a_L\leq n}S_{a_1}\cdots S_{a_L}.
\label{Llegop}
\end{equation}
In the loop gas picture, they have $L$ legs of open lines attached.
They are called the blobbed and unblobbed $L$-leg operators
\cite{Jacobsen:2006bn}.
They were named after the underlying Temperley-Lieb algebra though
we will not need its detailed property in this paper.
One of the important characteristics of these operators is that the
lines from blobbed operators can touch the JS boundary whereas
the line from unblobbed operators cannot.
We will give the matrix model equivalent of these operators
in the next subsection.

\subsection{The $O(n)$ matrix model}\label{sec:Onmm}

The $O(n)$ matrix model is an integral over $N\times N$ hermitian
matrices $X$ and $Y_a$, with $a$ running from 1 to $n$.
The partition function is given by \cite{Kostov:1988fy}
\begin{equation}
 Z=\int dX\prod_{a=1}^ndY_a
 \exp\left[ \beta\tr\Big(
  -\frac12X^2+\frac13X^3
  -\frac T2\sum_{a=1}^n Y_a^2+\sum_{a=1}^n XY_a^2 \Big)\right]\,.
\label{ZOnMM}
\end{equation}
The Feynman graph expansion of $Z$ generates all the
dynamical lattices of arbitrary genus but without boundaries.
The bare cosmological constant $\kappa$ is given by
\begin{equation}
 \beta=N\kappa^2,
\end{equation}
and each loop formed by the propagators of $Y_a$ is
multiplied by $nT^{-(\text{length})}$.
Graphs of genus $h$ are weighted by $N^{2-2h}$, so that
the planar graphs dominate the partition function
in the large $N$ limit for a fixed $\kappa$.
Continuum limit is obtained by sending $\kappa\to\kappa^*(T)$
and $N\to\infty$ in a correlated manner.

The physics in the continuum limit depends on the temperature.
Below the critical temperature $T<T^*$, partition function
is dominated by graphs with densely packed loops.
Since the vertices with three legs of $X$ do not play any role,
one could study the dense phase using the definition (\ref{ZOnMM})
without the $X^3$ term.
Before proceeding, we make a slight redefinitions of the matrices
$X$ and $Y_a$ so as to simplify the integrand of (\ref{ZOnMM}),
\begin{equation}
 Z=\int dX\prod_{a=1}^ndY_a
 \exp\left[-\beta\tr\Big(V(X)-\sum_{a=1}^nXY_a^2\Big)\right]\,.
\end{equation}
The potential $V(X)$ is then given by
\begin{equation}
V(X)=\frac12\big(X+\frac T2)^2 -\frac13\big(X+\frac T2\big)^3.
\label{defpot}
\end{equation}
As was mentioned above, generic quadratic potential $V(X)$ could
capture the physics in the dense phase.

The disc partition function with Neumann boundary condition
is given in the large $N$ limit by
\begin{equation}
\Phi(x)=-\frac1\beta\langle\tr\log(x-X)\rangle.
\end{equation}
Because of the prefactor $1/\beta$,
the leading contribution is independent of $N$ and
the higher genus contributions are subleading at large $N$.
What will become more important later is its derivative,
the resolvent
\begin{equation}
W(x)=-\frac{\partial}{\partial x}\Phi(x)
=\frac1\beta\Big\langle\tr\frac{1}{x-X}\Big\rangle\,.
\label{OnRes}
\end{equation}
The derivative introduces one marked point along the boundary.
One is supposed to take $x\to x^*=0$ in the continuum limit
\cite{Kostov:2006ry}.
The disc partition function with the $k$-th JS boundary condition
and one marked point is given by
\begin{equation}
\tilde W(y)=
\frac1\beta\Big\langle\tr\frac{1}{y-\sum_{a=1}^k{Y_a^2}}\Big\rangle\,,
\end{equation}
where we suppress the $k$-dependence of $\tilde W$
for notational simplicity.
To study the boundary changing operators, we also introduce
disc two-point functions with one Neumann
and one JS boundary conditions,
\begin{equation}
D_0(x,y)=\frac1\beta
   \Big\langle\tr\Big(\frac{1}{x-X}\frac{1}{y-\sum_{a=1}^k{Y_a^2}}
   \Big)\Big\rangle\,.
\label{defD0}
\end{equation}
We also consider the correlation functions of the $L$-leg operators,
\begin{align}
\begin{split}
D_L^\parallel(x,y) = \displaystyle\frac1\beta\Big\langle\tr\Big(
 \frac{1}{x-X}\mathbb{Y}_L^\parallel
 \frac{1}{y-\sum_{a=1}^k{Y_a^2}}\mathbb{Y}_L^\parallel \Big)\Big\rangle\,, \\
D_L^\perp(x,y) = \displaystyle\frac1\beta\Big\langle\tr\Big(
 \frac{1}{x-X}\mathbb{Y}_L^\perp
 \frac{1}{y-\sum_{a=1}^kY_a^2}\mathbb{Y}_L^\perp\Big)\Big\rangle\,,
\end{split}
\label{defDLOn}
\end{align}
where the operators $\mathbb{Y}_L^\parallel$ and
$\mathbb{Y}_L^\perp$ are defined analogously to (\ref{Llegop}),
\begin{equation}
 \mathbb{Y}_L^\parallel=
 \sum_{\{a_1,\cdots,a_L\}\subset\{1,\cdots,k\}}Y_{a_1}\cdots Y_{a_L},
 \qquad
 \mathbb{Y}_L^\perp=
 \sum_{\{a_1,\cdots,a_L\}\subset\{k+1,\cdots,n\}}Y_{a_1}\cdots Y_{a_L}.
\end{equation}
The sums are taken over all different sets of $L$ letters
(so $\mathbb{Y}_L^\parallel$ consists of $k!/L!$ terms).
These operators are the analogues in matrix model of the
blobbed and unblobbed $L$-leg operators.

In the following, we study the loop equations for the above
correlators that are associated to $Y_a$-derivatives.
We will see that the $Y_a$-derivative adds or removes one open line
between boundary operators, so that the loop equations relate the
correlators $D_L^{\perp,\parallel}$ with  $D_{L+1}^{\perp,\parallel}$.
Among those correlators, $D_1^\parallel$ and $D_0$ will be of
particular importance because they can be determined from a closed
system of shift equations.

\subsection{Loop equations}

We start from the loop equation which follows from the translation
invariance of the measure $dY_a$.
For any matrix $F$ made of $X$ and $Y_a$, the following equality holds:
\begin{equation}
\frac1\beta\sum_{ij}\Big\langle
\frac\partial{\partial Y_{aij}}F_{ij}\Big\rangle
 =-\big\langle\tr (FX+XF)Y_a \big\rangle.
\end{equation}
If we introduce $G=-(XF+FX)$, then $F$ is formally expressed
in terms of $G$ as
\begin{equation}
 F=\int_0^\infty d\ell e^{\ell X}Ge^{\ell X},
\end{equation}
Using them, the loop equation can be rewritten as
\begin{equation}
\sum_{ij}\frac1\beta
 \Big\langle \frac{\partial}{\partial Y_{aij}}\int_0^\infty d\ell
 (e^{\ell X}Ge^{\ell X})_{ij}
 \Big\rangle =\langle \tr\, GY_a \rangle.
\label{le}
\end{equation}
In the following we will apply this central identity
to different $G$ and derive some relations among our correlators
$D_L^\parallel$ and $D_L^\perp$.

\subsubsection{Loop equations for $D_0$ and $D_1^\parallel$}

For later convenience, we begin by introducing the notation
\begin{equation}
H(y)=\frac{1}{y-\sum_{a=1}^k{Y^2_a}}\,.
\end{equation}
Let us first apply (\ref{le}) to $G=e^{\ell'X}HY_a$ with $a\le k$.
Using the well known large $N$ factorization
$\langle\tr A\,\tr B\rangle\simeq
 \langle\tr A\rangle\langle\tr B\rangle$
and dropping the terms containing odd powers of $Y_a$ in a correlator,
we find
\begin{equation}
\beta\langle\tr\, e^{\ell'X}HY_a^2\rangle \;=\;
 \int d\ell
 \langle\tr\, e^{(\ell+\ell')X}H\rangle
 \Big(
 \langle\tr\, e^{\ell X}Y_aHY_a \rangle
+\langle\tr\, e^{\ell X}\rangle
 \Big).
\label{le1}
\end{equation}
Another relation can be obtained by applying (\ref{le})
to $G=e^{\ell'X}Y_aH$:
\begin{equation}
\beta\langle\tr\, e^{\ell'X}Y_aHY_a\rangle\;=\; 
\int d\ell \left(
  \langle\tr\, e^{(\ell+\ell')X}Y_aHY_a\rangle
 +\langle\tr\, e^{(\ell+\ell')X}\rangle
\right)
 \langle\tr\, e^{\ell X}H\rangle.
\label{le2}
\end{equation}
Now we make a Laplace transform with respect to $\ell$ and $\ell'$,
using the relations
\begin{align}
\begin{split}
& \int_0^\infty d\ell e^{-\ell x}\tr(e^{\ell X}A) =
 \tr\Big(\frac{1}{x-X}A\Big)\,, \\
& \int_0^\infty d\ell d\ell'
  e^{-x\ell'}\tr(e^{(\ell+\ell')X}A)\tr(e^{\ell X}B) =
 \tr\Big(\frac{1}{x-X}A\Big)\ast\tr\Big(\frac{1}{x-X}B\Big)\,.
\end{split}
\end{align}
Here we denoted by $\ast$ the Laplace transform of the convolution
\begin{equation}
F(x)\ast G(x)=\oint \frac{dx'}{2\pi i}\frac{F(x')-F(x)}{x-x'}G(-x')\,,
\label{defstar}
\end{equation}
and the contour of $x'$ integration here encircles around the cut
where $F(x)$ has discontinuity.
The loop equations (\ref{le1}) and (\ref{le2}) can then be rewritten
into the form
\begin{align}
\begin{split}
& \Big\langle\tr\Big( HY_a^2\frac{1}{x-X}\Big)\Big\rangle =
  D_0(x,y)\ast\left(
  \Big\langle\tr\Big( Y_aHY_a\frac{1}{x-X}\Big)\Big\rangle
 +\beta W(x) \right)\,,\\
& \Big\langle\tr\Big( Y_aHY_a\frac{1}{x-X}\Big)\Big\rangle =
  \left(\Big\langle\tr\Big( Y_aHY_a\frac{1}{x-X}\Big)\Big\rangle
 +\beta W(x)\right) \ast D_0(x,y)\,.
\end{split}
\end{align}
It only remains to sum over $a$ from $1$ to $k$.
Using the definitions (\ref{defD0}) and (\ref{defDLOn})
for $D_0$ and $D_1^\parallel$ as well as the equality
\begin{equation}
 \frac1\beta\sum_{a=1}^k
 \Big\langle\tr\Big(HY_a^2\frac{1}{x-X}\Big)\Big\rangle~=~
 yD_0(x,y)-W(x),
\end{equation}
we obtain
\begin{align}
\begin{split}
& yD_0(x,y)-W(x) = D_0(x,y)\ast\left(D_1^\parallel(x,y)+kW(x)\right), \\
&  D_1^\parallel(x,y) = \left(kW(x)+D_1^\parallel(x,y)\right)\ast D_0(x,y).
\end{split}
\label{leOn}
\end{align}

To obtain more useful equations for $D_0$ and $D_1^\parallel$,
we subtract their non-critical parts and define
\begin{align}
\begin{split}
& d_0(x,y)=D_0(x,y)-1, \\
& kd_1(x,y)=D_1^\parallel(x,y)+kW(x)-y.
\end{split}
\label{defOn}
\end{align}
It is natural to assume that $d_0$ and $d_1$ have
the same cut in the $x$-plane as that of $W(x)$,
since the cut is determined by the eigenvalue distribution of $X$
and therefore does not depend on the correlators considered.
The loop equation (\ref{leOn}) can then be rewritten in terms of
the discontinuity along the cut,
\begin{align}
\begin{split}
&d_0(-x,y)\Disc\, d_1(x,y) + \Disc\, W(x)=0, \\
&d_1(-x,y)\Disc\, d_0(x,y) + \frac1k\Disc\, W(x)=0,
\end{split}
\label{leOn00}
\end{align}
where $\Disc f(x) \equiv f(x+i0)-f(x-i0)$.
Using this one can show that the following quantity has
no discontinuities and no poles in the complex $x$-plane,
\begin{equation}
P_{10}(x,y)=d_1(x,y)d_0(-x,y)+W(x)+\frac1kW(-x)-\frac yk .
\end{equation}
Moreover, its behaviour at large $x$ is found from (\ref{defOn}),
\begin{equation}
 d_0 = -1+{\cal O}\big(\frac1x\big), \qquad
 d_1 = -\frac yk+{\cal O}\big(\frac1x\big), \qquad
 P_{10} = {\cal O}\big(\frac1x\big).
\end{equation}
Therefore $P_{10}$ should be identically zero.
\begin{align}
\begin{split}
&d_1(x,y)d_0(-x,y)+ W(x)+\frac1k W(-x)-\frac yk =0,\\
&d_0(x,y)d_1(-x,y)+W(-x)+\frac1k W( x)-\frac yk =0.
\end{split}
\label{leOn0}
\end{align}

As compare to our loop equation (\ref{leOn00}),
the equation obtained in \cite{Kostov:2007jj} (equations 3.13 and 3.14)
has one additional term coresponding to the JS boundary
touching itself to break the disc into two pieces.
Including such a term may be reasonable from the standpoint
of the combinatorics because the JS boundary
has fractal dimension $1/g$. In the matrix model description, this
boundary has classical dimension one and the term is missing. This
discrepency comes from two different possibilities for defining the
boundary Liouville potential. Because of the symmetry 
$\Delta_{r,s}(g)=\Delta_{s,r}(1/g)$ between the dressed scaling dimensions,
 both point of view give the same scaling dimension for the boundary operators.

\subsubsection{Loop equations for $D_L^\perp$ and $D_L^\parallel$}

Let us next take $G=e^{\ell'X}Y_aH$ with $a>k$ and apply (\ref{le}).
Following the similar steps as in the previous subsubsection, we get
\begin{equation}
 D_1^\perp(x,y)=(n-k)W(x)\ast D_0(x,y).
\end{equation}
This is actually a special case of more general recursion relations
between $D_{L+1}^\perp$ and $D_L^\perp$, and similarly between
$D_{L+1}^\parallel$ and $D_L^\parallel$.
To derive them, let us denote by $\{a_i\}, \{b_i\}$ two arbitrary
sets of order $L+1$.
They are both chosen to be subsets of $\{1,\cdots,k\}$
or $\{k+1,\cdots,n\}$ depending on whether we are interested
in $D_L^\parallel$ ot $D_L^\perp$.
Then we apply the loop equation (\ref{le}) to
\begin{equation}
G\;=\;Y_{a_L}\cdots Y_{a_1}
 \frac{1}{y-\sum_{c=1}^k{Y_c^2}}Y_{b_1}\cdots Y_{b_{L+1}}e^{\ell'X},
\end{equation}
where the derivative is with respect to $Y_{a_{L+1}}$,
and sum over the sets $\{a_i\}$ and $\{b_i\}$.
The final result is
\begin{align}
\begin{split}
& D_{L+1}^\parallel(x,y) = (k-L)W(x)\ast D_L^\parallel(x,y), \\
& D_{L+1}^\perp(x,y) = (n-k-L)W(x)\ast D_L^\perp(x,y).
\end{split}
\label{leOnL}
\end{align}
These relations agree with the result of \cite{Kostov:2007jj}.
In terms of discontinuity along the cut, the loop equations become
\begin{align}
\begin{split}
&\Disc\, D_{L+1}^\parallel(x,y)=(k-L) D_L^\parallel(-x,y) \Disc\, W(x), \\
&\Disc\, D_{L+1}^\perp(x,y)=(n-k-L) D_L^\perp(-x,y) \Disc\, W(x).
\end{split}
\label{leOnL2}
\end{align}
The second equation can be extended to the case $L=0$ if one defines
\begin{equation}
 D_0^\perp(x,y)=D_0(x,y), \qquad
 D_0^\parallel(x,y)=\frac{D_0}{1-D_0}.
\end{equation}
It was pointed out in \cite{Kostov:2007jj} that these relations
follow naturally if the Neumann and JS boundaries are allowed to
touch in $D_0^\parallel$ but not allowed in $D_0^\perp$.

\subsection{Solution in the continuum limit}

\subsubsection{The disc amplitude $W$}

To study the continuum limit, we introduce a small parameter $\epsilon$
and set the unit lattice length to $\epsilon$.
We define the renormalized bulk and boundary cosmological constants
$(\mu,\xi,\zeta)$ by \cite{Kostov:1991cg}
\begin{equation}
 \epsilon^2\mu = \kappa-\kappa^*,\qquad
 \epsilon^{1/g}\xi = x-x^*, \qquad
 \epsilon\zeta = y-y^*.
\end{equation}
They blow up the neighbourhood of the critical point
$(\kappa^*,x^*, y^*)$ in the scaling limit $\epsilon\to 0$.
Note that the JS boundaries have classical dimension $1$
whereas the Neumann boundary has fractal dimension $1/g$.
The renormalized resolvent $w(\xi)$ is defined by
\begin{equation}
W(x)-\frac{2V'(x)-nV'(-x)}{4-n^2}=\epsilon w(\xi).
\label{renorW}
\end{equation}
In \cite{Kostov:1991cg} the resolvent was
obtained in the following parametric form,
\begin{equation}
\xi=M\cosh\tau, \qquad
w(\xi)=-\frac{M^g}{2g}\cosh g\tau.
\label{wxitau}
\end{equation}
Here $M$ is related to the cosmological constant $\mu$ and
the string susceptibility $u$ via
\begin{equation}
  M^{2g}=2g\mu,\qquad
  u=\frac{\partial^2 Z_\text{sphere}}{\partial\mu^2}=M^{2\theta}.
\label{Mmuu}
\end{equation}
As a function of $\xi$, the resolvent has a cut along the interval
$\,]\!-\!\infty,-M]$ where the eigenvalues of the matrix $X$ are distributed.

\subsubsection{The disc amplitudes $D_0^\perp$ and $D_1^\parallel$}

Here we wish to solve (\ref{leOn0}) in the continuum limit.
To begin with, we need to find the critical value $y^*$ of
the $k$-th JS boundary cosmological constant.
Since it should not depend on $x$ and $\kappa$,
one can determine it by requiring that $d_0$ and $d_1$ vanish at
the critical point $(x,y,\kappa)=(0,y^*,\kappa^*)$,
\begin{equation}
y^*=(k+1)W(0).
\label{yOn}
\end{equation}
With a slight abuse of notations, we define
the renormalized two-point functions by
\begin{equation}
d_i(x,y)=\epsilon^{\alpha_i/g}d_i(\xi,\zeta),
\end{equation}
where the scaling exponents $\alpha_0$, $\alpha_1$
will be determined shortly.
To the leading order in small $\epsilon$,
the loop equations (\ref{leOn0}) become
\begin{align}
\begin{split}
& d_1(\xi,\zeta)d_0(-\xi,\zeta)+w( \xi)+\frac1k w(-\xi)-\frac\zeta k = 0,\\
& d_0(\xi,\zeta)d_1(-\xi,\zeta)+w(-\xi)+\frac1k w( \xi)-\frac\zeta k = 0.
\end{split}
\label{leOnc}
\end{align}
We dropped several terms in (\ref{leOn0}) such as polynomial terms
in $W(x)$ because they are subdominant for small $\epsilon$.
By noticing $d_i\sim \xi^{\alpha_i}$ for $\zeta=\mu=0$ and
using the loop equations, one can determine the scaling exponents
\cite{Kostov:2007jj},
\begin{equation}
 \alpha_0=r\theta, \qquad
 \alpha_1=1-\theta-r\theta,
\label{a0a1}
\end{equation}
where $r$ is related to $k$ by
\begin{equation}
k(r)=\frac{\sin{(r+1)\pi\theta}}{\sin{r\pi\theta}}.
\label{defk}
\end{equation}
Hereafter we use $r$ as the label of JS boundaries;
it has a clear physical meaning as we will see later.
We also express $\zeta$ in terms of a new parameter $\sigma$ as
\begin{equation}
\zeta(\sigma)=
\frac{M^g}{2g}\frac{\sin\pi\theta}{\sin\pi r\theta}\cosh g\sigma.
\label{zetas}
\end{equation}

Using (\ref{wxitau}), (\ref{defk}) and (\ref{zetas})
the loop equations can be rewritten as
\begin{equation}
   d_1\big(\tau\mp\frac{i\pi}2,\sigma\big)
   d_0\big(\tau\pm\frac{i\pi}2,\sigma\big)
  = CM^g\cosh\frac{g(\tau+\sigma)\pm\alpha}{2}
        \cosh\frac{g(\tau-\sigma)\pm\alpha}{2},
\label{d0d1le}
\end{equation}
where
\begin{equation}
C=\frac{\sin\pi\theta}{g\sin\pi(r+1)\theta},\qquad
\alpha =\frac{i\pi}{2}(2r\theta+\theta-1).
\end{equation}
We solve these shift relations in Appendix \ref{sec:sol}
using a slight generalization of \cite{Kazakov:1991pt}.
The result is that $d_0, d_1$ are given by the Liouville boundary
two-point function \cite{Fateev:2000ik}.
See Appendix \ref{sec:kpz} for its explicit form.
The scaling exponents of the correlators $d_i\propto\xi^{\alpha_i}$
are then read from their dependence on the boundary cosmological constant,
and agree precisely with (\ref{a0a1}).
This is enough to determine the gravitational dimensions of
the boundary changing operators in the correlators $d_0$ and $d_1$,
\begin{equation}
 \Delta_0^\perp=\Delta_{r,r}, \qquad
 \Delta_1^\parallel=\Delta_{-r,-r-1},
\end{equation}
where the gravitational dimension for the $(r,s)$ operator
is given by
\begin{equation}
 \Delta_{r,s}=\frac{r-1-g(s-1)}{2g},
\end{equation}
and is related to the conformal weight of the operators in CFT by
KPZ relation (\ref{EquKPZ}).
See Appendix \ref{sec:kpz} for more detail.

\subsubsection{The disc amplitudes $D_L^\perp$ and $D_L^\parallel$}

Using the parameters $(\tau,\sigma)$ in the continuum limit and
the equality
\begin{equation}
w(\tau+i\pi)-w(\tau-i\pi)= \frac{M^g}{ig}\sin\pi\theta\sinh g\tau,
\end{equation}
the loop equations (\ref{leOnL2}) can be rewritten into the form
\begin{align}
\begin{split}
& D_{L+1}^\parallel(\tau+i\pi,\sigma)
 -D_{L+1}^\parallel(\tau-i\pi,\sigma)
 =\frac{(k-L)M^g}{ig}\sin\pi\theta\sinh g\tau
  D_L^\parallel(\tau,\sigma),\\
& D_{L+1}^\perp(\tau+i\pi,\sigma)
 -D_{L+1}^\perp(\tau-i\pi,\sigma)
 =\frac{(n-k-L)M^g}{ig}\sin\pi\theta\sinh g\tau D_L^\perp(\tau,\sigma).
\end{split}
\end{align}
These shift relations take the same form as (\ref{shiftFZZ})
satisfied by the Liouville boundary two-point function.
So, in the continuum limit the correlators $D_L^\parallel$ and
$D_L^\perp$ are again given by Liouville boundary two-point functions
up to factors independent of $\tau,\sigma$.
The gravitational dimensions of the $L$-legs boundary operators
are given by
\begin{equation}
 \Delta_L^\parallel=\Delta_{-r,-r-L}, \qquad
 \Delta_L^\perp=\Delta_{r,r-L}.
\end{equation}
Through KPZ relation (\ref{EquKPZ}) this determines
the conformal weight of the blobbed and unblobbed operators.
The results are in complete agreement with \cite{Jacobsen:2006bn}
and confirm our identification of matrix model correlators
with those of the $O(n)$ model coupled to gravity.

As was noticed in \cite{Jacobsen:2006bn}, there is no reason
for the parameter $r$ to be quantized in the $O(n)$ model.
So we have a continuous spectrum of JS boundary conditions
and the associated boundary-changing operators.
We conclude this section with a few remarks on some special values of $r$.
First, for $k=n$ or $r=1$ the JS boundary becomes the Neumann boundary.
In this case the conformal weights of the boundary operators become
$\delta_L^\perp=\delta_{1,L+1}$ and $\delta_L^\parallel=\delta_{1,L+1}$,
in agreement with the result \cite{Saleur:1987rn} on flat lattice
and \cite{Kazakov:1991pt} on dynamical lattice.
Another interesting special case is the Dirichlet case,
which corresponds to $k=1$ or
\begin{equation}
r=\frac{1-\theta}{2\theta}.
\end{equation}
In this limit, using the fact that the correlators of odd powers
of $Y_a$ matrices vanish, one can prove the relations
\begin{align}
\begin{split}
& D_0^\perp(x,y)
 =\Big\langle\tr\frac{1}{x-X}\frac{1}{y-Y_a^2}\Big\rangle
 =\frac1{\sqrt y}
  \Big\langle\tr\frac{1}{x-X}\frac{1}{\sqrt y-Y_a}\Big\rangle,\\
& D_1^\parallel(x,y)
 =yD_0^\perp(x,y)-W(x)
 =\sqrt y\Big\langle\tr\frac{1}{x-X}\frac{1}{\sqrt{y}-Y_a}\Big\rangle-W(x).
\end{split}
\end{align}
The loop equation for $k=1$ then involves only one undetermined quantity,
\begin{equation}
yd_0(x,y)d_0(-x,y)+W(x)+W(-x)-y=0.
\end{equation}
We recovered the loop equation for the correlation functions
of twist operators in loop gas model \cite{Kazakov:1991pt}
up to normalization of correlators and parameters.
Note that this simplification is a special feature of $k=1$
because the resolvents for the JS boundaries with $k>1$ involve
$k$ non-commuting matrices $Y_a$.

\section{Boundary operators in RSOS model}\label{sec:sos}

\subsection{Definition of the model}

In the RSOS height model \cite{Andrews:1984af},
the local fluctuation variable (height) takes values
in the integer set $\{1,\cdots,h-1\}$.
This model is also called $A_{h-1}$-model,
the integer set being identified with the nodes
of the Dynkin graph for the $A$ series.
This graph is characterized by its adjacency matrix
\begin{equation}
G_{ab}=\begin{cases}
1& \text{if}~ a\sim b\,,  \cr
0& \text{otherwise}.    
\end{cases}
\end{equation}
The indices $a,b$ run over the nodes of the Dynkin graph,
and $a\sim b$ means that the nodes $a,b$ are linked.
One can define the so called ADE-models in the same way from
the Dynkin graphs of the ADE Lie algebras \cite{Pasquier:1986jc}.

The RSOS model on a fixed triangular lattice with possible
curvature defects is defined as the statistical sum over all
the height configurations.
Each height configuration is weighted according to
the following rule \cite{Kostov:1995xw}.
To each site of height $a$ one assigns the local Boltzmann weight
\begin{equation}
 W_\circ(a)=S_a,
\label{wsite}
\end{equation}
and to each triangle with the heights $a,b,c$ at the three vertices
one assigns
\begin{equation}
 W_\Delta(a,b,c)=\frac{1}{\sqrt{S_a}}\delta_{ab}\delta_{bc}+
  \frac{1}{T\sqrt{S_a}}\Big(
      \delta_{ab}G_{bc}+\delta_{bc}G_{ca}+\delta_{ca}G_{ab}\Big).
\label{wtrig}
\end{equation}
Here $T$ is the temperature, and $S_a$ are the components of
the Perron-Frobenius vector
\begin{equation}
S_a=\sqrt{\frac2h}\sin{\left(\frac{\pi a}{h}\right)},
\label{eIiv}
\end{equation}
which is the eigenvector of the adjacency matrix
with the largest eigenvalue $2\cos{(\pi/h)}$.
The weight (\ref{wtrig}) in particular requires that
the heights of any two adjacent sites can differ at most by a unit.
Thanks to this, the height configurations can also be described
by the contour lines (loops) along the edges of the dual lattice
\cite{Nienhuis:1984wm}.
The average total length of the loop is
then controlled by the temperature.

The phase diagram of the RSOS model is
the same as that for the $O(n)$ model;
it is in the dilute phase at some critical temperature $T^*$,
and in the dense phase at lower temperatures.
Since the lattice is filled by the loops in the dense phase,
one can study this phase using the Boltzmann weight (\ref{wtrig})
without the first term.

Conformal boundary conditions are realized microscopically
as suitable restrictions on the heights on the boundary.
In the dense phase there are two kinds of boundary conditions.
The ``fixed'' boundary condition $a$ requires the boundary
to take the constant height $a$.
The ``alternating'' boundary conditions $\link{ab}$ require
the sites at the boundary to take the two adjacent heights
$a$ and $b$ alternately, like $ababab...ab$.
Bauer and Saleur \cite{Saleur:1988zx} studied both types of boundary conditions
on the flat lattice, and found that conformal weights of
the boundary changing operators ${}^1\!B{}^a$ and
${}^{\link{12}}\!B{}^{\link{ab}}$ are given by $\delta_{1,a}$
and $\delta_{a,1}$.
In the following we will show that the boundary operators of
conformal weight $\delta_{r,s}$ can be realized as
${}^s\!B{}^{\link{r\,r+1}}$ when $r\ge s$ and
${}^s\!B{}^{\link{r+1\,r}}$ when $r<s$.

\subsection{The RSOS matrix model}

The RSOS matrix models \cite{Kostov:1992ie} are the simplest examples
of the $ADE$-matrix models.
The fluctuating variables are $h-1$ hermitian matrices $X_a$
associated to the nodes $a=1,\cdots,h-1$,
and $h-2$ rectangular complex matrices $C_{ab} =C^\dagger_{ba}$
associated with the oriented links $\link{ab}$.
$X_a$ has the size $N_a\times N_a$ while $C_{ab}$ has the size
$N_a\times N_b$.
The partition function is given by the integral
\begin{align}
\begin{split}
& Z=\int\prod_adX_a\prod_{\link{ab}}dC_{ab}
  \exp\big(-\beta S[X,C]\big), \\
& S[X,C]=\sum_a S_a\tr\left(\frac12X_a^2-\frac13X_a^3\right)
        +\sum_{\link{ab}}\tr\left(\frac T2 C_{ab}C_{ba}
                     -X_aC_{ab}C_{ba}\right),
\end{split}
\end{align}
where $\link{ab}$ runs over oriented links of the Dynkin graph.

We take the limit of large $\beta$, large $N_a$ keeping their
ratio fixed,
\begin{equation}
 \frac{N_a}{\beta S_a}\equiv\kappa^2= \text{fixed}.
\label{defb}
\end{equation}
The constant $\kappa$ plays the role of the bare cosmological constant.
Perturbative expansion of $Z$ gives the sum
over height configurations with Boltzmann weights
(\ref{wsite}) and (\ref{wtrig}), but now on dynamical lattice.
The planar graphs dominate the partition function
in the large $\beta$ limit, and the higher genus terms are
suppressed by powers of $\beta^{-2}$.

One can view the system as the gas of loops on dynamical lattice
which are formed by the propagators of $C_{ab}, C_{ba}$
and separating the domains of heights $a$ and $b$.
The temperature $T$ is regarded as the fugacity for the total loop length.
Again, the term $X_a^3$ in $S[X,C]$ can be dropped when one is interested
in the dense phase.

Also, the Feynman rule is such that each connected domain
of height $a$ gives rise to a factor
\begin{equation}
 (S_a)^\chi,
\label{flSOS}
\end{equation}
where $\chi$ is the Euler number of that domain.
For disc graphs, this rule amounts to assigning the factor
$S_a$ to each ``outermost'' domain of height $a$ touching the boundary, and
$S_a/S_b$ to each domain of height $a$ surrounded by that of height $b$.

Before proceeding, we make a certain linear change of matrix variables
to rewrite the partition function as
\begin{equation}
 Z = \int\prod_a dX_a\prod_{\link{ab}}dC_{ab}
   \exp\Big(-\beta\sum_aS_a\tr V(X_a)
  +\beta\sum_{\link{ab}}X_aC_{ab}C_{ba}\Big),
\end{equation}
where the potential $V(x)$ is the same as the one (\ref{defpot})
for the $O(n)$ model.
As was mentioned above, generic quadratic $V$ can describe the
physics in the dense phase.

We will consider the following correlators
\begin{align}
\begin{split}
&\Phi_a(x)=-\frac1\beta\big\langle\tr\log(x-X_a)\big\rangle, \\
&\Phi_{\link{ab}}(y)=-\frac1\beta\big\langle\tr\log(y-C_{ab}C_{ba})\big\rangle.
\end{split}
\end{align}
They correspond respectively to the disc partition function with
the boundary conditions $a$ or $\link{ab}$.
The more important quantities in the following analysis are
their derivatives, the loop amplitudes
\begin{align}
\begin{split}
& W_a(x)  = -\frac{\partial}{\partial x}\Phi_a(x)
          =\frac1\beta\Big\langle\tr\,\frac{1}{x-X_a}\Big\rangle\,,\\
& W_{\link{ab}}(y) = -\frac{\partial}{\partial y}\Phi_{\link{ab}}(y)
          =\frac1\beta\Big\langle\tr\,\frac{1}{y-C_{ab}C_{ba}}\Big\rangle\,.
\end{split}
\label{defWa}
\end{align}
The loop amplitudes $W_a(x)$ are the Laplace images of the loop
amplitudes with fixed boundary length,
\begin{equation}
 \tilde W_a(\ell) = \frac1\beta\big\langle\tr\, e^{\ell X_a}\big\rangle\, .
\label{defWl}
\end{equation}
To study the spectrum of boundary operators, we also consider
the disc two-point functions with $a$ and $\link{bc}$ boundary
segments,
\begin{equation}
 D_{a,\link{bc}}(x,y)=\frac1\beta\Big\langle\tr\Big(
 \frac{1}{x-X_a}{\bf S}^L_{ab}\frac{1}{y-C_{bc}C_{cb}}{\bf S}^L_{ba}
 \Big)\Big\rangle\,.
\label{DefDL}
\end{equation}
Here 
\begin{equation}
 {\bf S}^L_{ab} \equiv C_{a d_1}C_{d_1d_2}\cdots C_{d_{L-1} b}
\end{equation}
is the product of $C$'s along the shortest path from $a$ to $b$,
so that each height $d_i$ appears only once.
We also assume that the path does not contain the node $c$.
There are $L$ open contour lines stretching between
the two boundary changing operators ${\bf S}^L_{ab}$.
Their Laplace transform,
\begin{equation}
\tilde D_{a,\link{bc}}(\ell,y)
=\frac1\beta \Big\langle\tr\Big( e^{\ell X_a}
 {\bf S}^L_{ab}\,\frac{1}{y-C_{bc}C_{cb}}{\bf S}^L_{ba}\Big)\Big\rangle\,,
\label{Dabcl}
\end{equation}
will also appear in the derivation of the loop equation.

\subsection{Loop equations}

Here we derive some loop equations for disc correlators by
using the same technique that was used in the $O(n)$ model.
Let us start from
\begin{equation}
 \frac1\beta\Big\langle
 \frac{\partial F_{ij}}{\partial C_{ab\;ij}}\Big\rangle
 =-\big\langle\tr\, F(C_{ba}X_a+X_bC_{ba})\big\rangle\,,
\end{equation}
where $F$ is a $N_a\times N_b$ matrix made of $X$ and $C$ matrices.
By inserting
\begin{equation}
F=\int_0^\infty{d\ell e^{\ell X_a}Ge^{\ell X_b}},
\end{equation}
we obtain the following equation
\begin{equation}
 \sum_{ij}\frac1\beta
 \Big\langle\frac{\partial}{\partial C_{ab\;ij}}
 \int_0^\infty dl\big(e^{\ell X_a}G e^{\ell X_b}\big)_{ij}\Big\rangle
 = \big\langle\tr\, C_{ba}G \big\rangle\,.
\label{Rle}
\end{equation}

\subsubsection{Recursion relation for $D_{a,\link{bc}}$}

To begin with, we derive a loop equation involving $D_{a,\link{bc}}$
assuming $a\ne b$ and that $c$ is not on the shortest path connecting
$a$ and $b$.
Let $d$ be the node adjacent to $a$ along that shortest path,
so that ${\bf S}^L_{ab}=C_{ad}{\bf S}^{L-1}_{db}$ and
${\bf S}^L_{ba}={\bf S}^{L-1}_{bd}C_{da}$.
Applying (\ref{Rle}) to
\begin{equation}
G=e^{\ell'X_a}C_{ad}\,{\bf S}^{L-1}_{db}\frac{1}{y-C_{bc}C_{cb}}
                    {\bf S}^{L-1}_{bd}\,,
\end{equation}
gives the following recursion relation
\begin{equation}
 \tilde D_{a,\link{bc}}(\ell',y)
 =\int_0^\infty d\ell\,\tilde W_a(\ell+\ell')
  \tilde D_{d,\link{bc}}(\ell,y)\,.
\end{equation}
Its Laplace image reads
\begin{equation}
 D_{a,\link{bc}}(x,y) = W_a(x)\ast D_{d,\link{bc}}(x,y),
\label{recIntz}
\end{equation}
with the $\ast$ product defined in (\ref{defstar}).
In this derivation, it is important that $C_{ad}$ appears only once.
The relation (\ref{recIntz}) can also be obtained by an explicit
integration over the $C_{ad}$ matrix using the Gaussian measure.
From the viewpoint of height configuration, this can be understood
as cutting the disc into two pieces along the contour line
that stretches between the two boundary operators
and separates the domains of height $a$ and $d$.
See Figure \ref{fig:sos1}.
\FIG{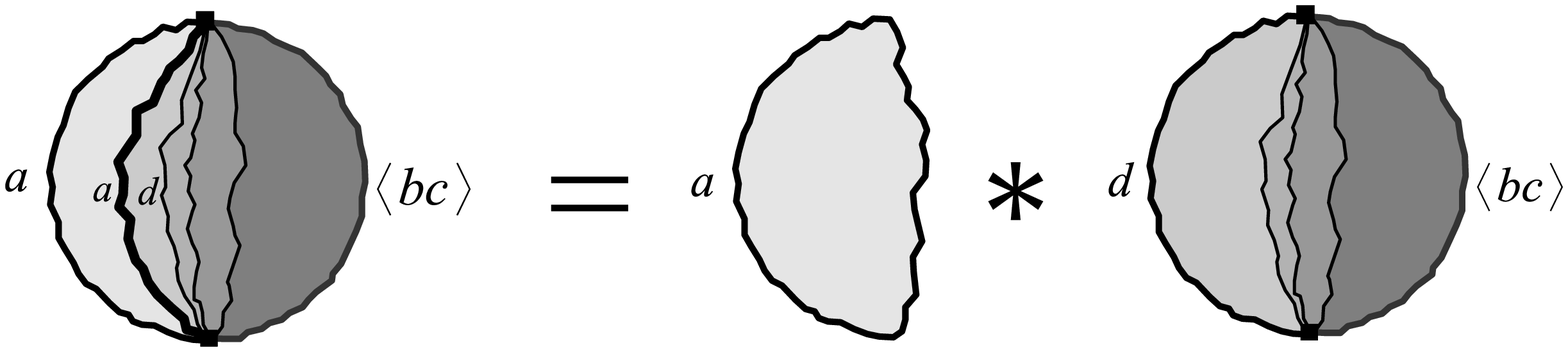}
    {The disc graph contributing to $D_{a,\link{bc}}$ can be decomposed into
     two discs along the contour line separating the domains of height
     $a$ and $d$ and connecting the two boundary operators.
     $D_{a,\link{bc}}$ is therefore written as a $\ast$-product of
     $W_a$ and $D_{d,\link{bc}}$.
    }{fig:sos1}
This recursion relation allows us to express all the disc correlators
$D_{a,\link{bc}}$ in terms of the $2h-2$ basic correlators
$D_{a,\link{a\,a\pm1}}$.

\subsubsection{Bilinear functional equation for $D_{a,\link{ab}}$}

Let us next apply (\ref{Rle}) to
\begin{equation}
G=e^{\ell'X_a}\frac{1}{y-C_{ab}C_{ba}}C_{ab}\,.
\end{equation}
Following the similar steps as before we obtain the loop equation
\begin{equation}
 y\int_0^\infty d\ell \tilde D_{a,\link{ab}}(\ell+\ell',y)
 \tilde D_{b,\link{ba}}(\ell,y) =
 y\tilde D_{a,\link{ab}}(\ell',y)-\tilde W_a(\ell').
\end{equation}
After the Laplace transform with respect to $\ell$,
it becomes
\begin{equation}
 D_{a,\link{ab}}(x,y)=\frac1y W_a(x)
+D_{a,\link{ab}}(x,y)\ast D_{b,\link{ba}}(x,y).
\label{Rleb}
\end{equation}
This relation can also be understood from the viewpoint of the
height configurations.
Recall first of all that the $\link{ab}$ boundary emanates
contour lines separating the heights $a$ and $b$.
Let us then think of a graph participating in $D_{a,\link{ab}}$,
and cut it along the contour line whose endpoint is
the closest to the right end of the $\link{ab}$ boundary.
There may be no such contour line because the $\link{ab}$
boundary may have zero length; in such a case the graph has
constant boundary height $a$ and contributes to $W_a(x)$.
Otherwise the graph can be decomposed into two pieces contributing
respectively to $D_{a,\link{ab}}$ and $D_{b,\link{ba}}$.
See Figure \ref{fig:sos2}.
\FIG{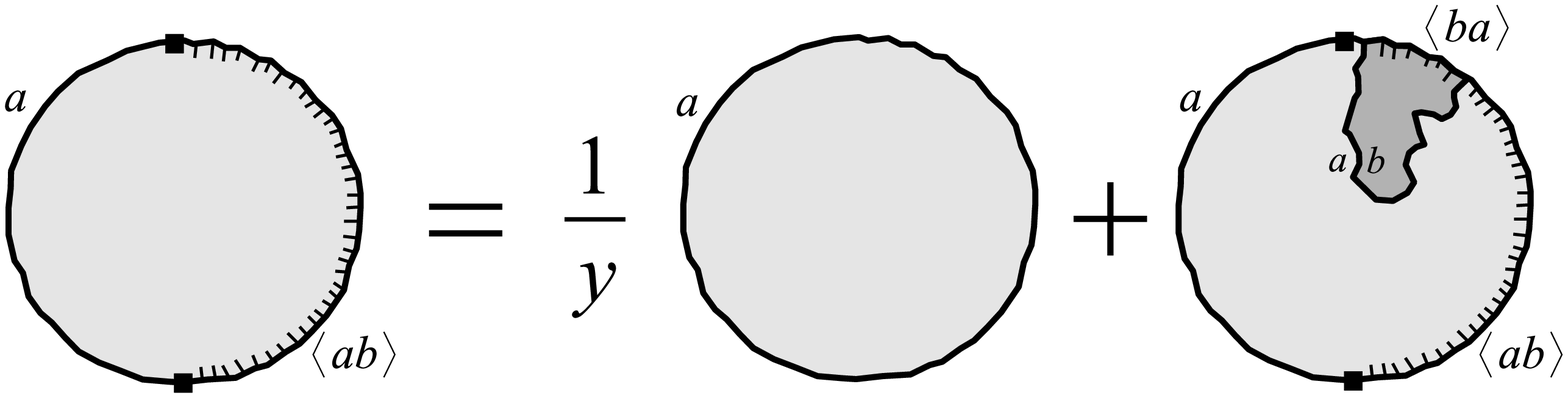}
    {There are two types of graphs contributing to $D_{a,\link{ab}}$.
     The graph may have $\link{ab}$ boundary of length zero so that the
     whole boundary has height $a$.
     If $\link{ab}$ boundary has nonzero length, one can cut it into two
     pieces contributing to $D_{a,\link{ab}}$ and $D_{b,\link{ba}}$,
     respectively.
    }{fig:sos2}
These two possibilities correspond to the two terms in the RHS of
(\ref{Rleb}).

The disc correlators $D_{a,\link{ab}}$ and $D_{b,\link{ba}}$
satisfy two relations, namely (\ref{Rleb}) and another equation
obtained by exchanging $a$ and $b$.
To get the more useful equation we take the discontinuity of
these equations along the cut in the $x$-plane,
\begin{align}
\begin{split}
 \Disc D_{a,\link{ab}}(x,y)=
 \frac1y\Disc W_a(x)+D_{b,\link{ba}}(-x,y)\Disc D_{a,\link{ab}}(x,y),\\
 \Disc D_{b,\link{ba}}(x,y)=
 \frac1y\Disc W_b(x)+D_{a,\link{ab}}(-x,y)\Disc D_{b,\link{ba}}(x,y).
\end{split}
\end{align}
Using these and taking into account the large-$x$ asymptotics
of $W_a$ and $D_{a,\link{ab}}$, we find
\begin{align}
&  yD_{a,\link{ab}}(x,y)D_{b,\link{ba}}(-x,y)  \nonumber \\
& -yD_{a,\link{ab}}(x,y)-yD_{b,\link{ba}}(-x,y)
  +W_a(x)+W_b(-x) ~=~ 0.
\label{leRSOS}
\end{align}
By introducing
\begin{equation}
d_{a,\link{ab}}(x,y)=\sqrt{y}\big(D_{a,\link{ab}}(x,y)-1\big)
\end{equation}
we can rewrite the equation (\ref{leRSOS}) into a simpler form,
\begin{equation}
d_{a,\link{ab}}(x,y)d_{b,\link{ba}}(-x,y)+W_a(x)+W_b(-x)=y.
\label{renorleRSOS}
\end{equation}
Here the normalization of $d_{a,\link{ab}}$ is somewhat arbitrary,
and we chose the symmetric normalization for simplicity.

\subsection{Solution in the continuum limit}

To study the continuum limit, we introduce a small parameter
$\epsilon$ and the renormalized couplings $(\mu,\xi,\zeta)$
in the same way as we did for the $O(n)$ model,
\begin{equation}
 \epsilon^2\mu = \kappa-\kappa^*,\qquad
 \epsilon^{1/g}\xi = x-x^*, \qquad
 \epsilon \zeta = y-y^*.
\end{equation}
In \cite{Kostov:1991cg,Kostov:1989eg} the loop equations for $W_a(x)$
has been solved under the natural ansatz,
\begin{equation}
W_a(x) = S_aW(x).
\label{waw}
\end{equation}
$W(x)$ was then shown to satisfy the loop equation
(\ref{OnRes}) for the resolvent of the $O(n)$ matrix model
with $n=2\cos(\pi/h)$.
So we borrow the solution from the $O(n)$ model
under the identification
\begin{equation}
  \theta=1-g=\frac 1h.
\end{equation}
We define the renormalized resolvent $w(\xi)$ in the same way
as in (\ref{renorW}).
The solution in the continuum limit is given by
\begin{equation}
\xi=M\cosh\tau, \qquad
w(\xi)=-\frac{M^g}{2g}\cosh g\tau.
\label{xiwtau2}
\end{equation}
Again, $M$ is related to $\mu$ via (\ref{Mmuu}).

Now let us solve the loop equation (\ref{renorleRSOS}) for
the correlators $d_{a,\link{ab}}$.
First we need to determine the critical value
of the $\link{ab}$-boundary cosmological constant $y^\ast$.
We require that $d_{a,\link{ab}}$ and $d_{b,\link{ba}}$ vanish
at the critical point $(\kappa^*,x^*,y^*)$ and find
\begin{equation}
y^*=W_a(0)+W_b(0)=(S_a+S_b)W(0).
\label{yRSOS}
\end{equation}
Next we renormalize the disc correlators near the critical point as
\begin{align}
\begin{split}
& d_{a,\link{ab}}(x,y)=\epsilon^{\alpha_{a,\link{ab}}/g}
  d_{a,\link{ab}}(\xi,\zeta), \\
& \alpha_{a,\link{ab}}+\alpha_{b,\link{ba}}=g=1-\frac1h\,.
\end{split}
\end{align}
The loop equation in the limit of small $\epsilon$ is
\begin{equation}
 d_{a,\link{ab}}(\xi,\zeta)
 d_{b,\link{ba}}(-\xi,\zeta)+S_aw(\xi)+S_bw(-\xi)=\zeta.
\label{renorleRSOS2}
\end{equation}
By substituting (\ref{xiwtau2}) and
\begin{equation}
\zeta=\frac{M^gS_1}{2g}\cosh{g\sigma}
\end{equation}
into (\ref{renorleRSOS2}), the loop equation finally becomes
\begin{align}
\begin{split}
& d_{a,\link{ab}}\big(\tau-i\pi/2,\sigma\big)
  d_{b,\link{ba}}\big(\tau+i\pi/2,\sigma\big)
 \\ &
 =\frac{M^gS_1}{g}
        \cosh\Big(\frac{g(\tau+\sigma)+\alpha_{ab}}{2}\Big)
        \cosh\Big(\frac{g(\tau-\sigma)+\alpha_{ab}}{2}\Big),
\end{split}
\end{align}
where $\alpha_{ab}$ for $b=a\pm1$ is given by
\begin{equation}
\alpha_{ab}=\pm\frac{i\pi}{2}\left(1-\frac{a+b}{h}\right)=-\alpha_{ba}.
\end{equation}

The above equation has the same structure as (\ref{d0d1le}),
so the solution is given in terms of the Liouville boundary
two-point functions.
We can now use the formulae in Appendix \ref{sec:kpz} and \ref{sec:sol}
and determine the gravitational dimensions of the boundary
changing operators
\begin{alignat}{3}
 \Delta_{a,\link{ab}}&=\Delta_{a,a},   \quad
&\Delta_{b,\link{ba}}&=\Delta_{1-b,-b} \quad
&\text{when }b=a+1, \nonumber \\
 \Delta_{a,\link{ab}}&=\Delta_{1-a,-a},\quad 
&\Delta_{b,\link{ba}}&=\Delta_{b,b}    \quad
&\text{when }b=a-1.
\end{alignat}
This implies that the boundary operators ${}^a\!B{}^{\link{a\,a+1}}$
and ${}^a\!B{}^{\link{a\,a-1}}$ in RSOS model have conformal weights
$\delta_{a,a}$ and $\delta_{a-1,a}$, in full agreement
with the result of Saleur and Bauer.

From the comparison of the loop equation (\ref{recIntz})
with (\ref{shiftFZZ}), it follows that the correlators $d_{a,\link{bc}}$
are all given by Liouville boundary two-point functions.
The operators ${}^{a-L}\!B{}^{\link{a\,a+1}}$ and
${}^{a+L}\!B{}^{\link{a\,a-1}}$ are then shown to have
conformal weights $\delta_{a,a+L}$ and $\delta_{a-1,a+L}$,
respectively.
By varying the integer parameters $a$ and $L$,
the whole spectrum of boundary operators for this rational CFT
is recovered.
This result proves the conjecture of \cite{Saleur:1988zx}
on the scaling dimensions of boundary changing operators
in the RSOS model.

\section{The map between the two models}\label{sec:map}

It has been known for a long time that the $O(n)$ model and SOS models
are described by the same class of conformal field theories.
In particular, the RSOS model with $h-1$ nodes was known to have
the same partition function on the plane as the rational $O(n)$ model
with $n=2\cos(\pi/h)$.
Here we wish to explore this correspondence further, focusing
mainly on the theories on the disc.

As was used in the previous section, the $O(n)$ model on the disc
with Neumann boundary condition is equivalent to the RSOS model
with fixed-height boundary condition.
The disc partition functions are related via
\begin{equation}
W_a(x_\text{RSOS})=S_aW(x_\text{O(n)}).
\label{waw2}
\end{equation}
In \cite{Jacobsen:2006bn} it was shown that the $\link{ab}$-type boundaries
of RSOS model correspond to the JS boundaries of the $O(n)$ model
labeled by integer $r$.
The annulus partition functions of the $O(n)$ model with
Neumann-JS boundary conditions were shown to agree with those
of RSOS model with $a$-$\link{bc}$ boundary conditions.
It was also noticed there that one needs to introduce
$L$ non-contractible loops on the annulus of the $O(n)$ model,
corresponding to the distance between $a$
and $\link{bc}$ labelling the two boundaries
of the RSOS model.

Following their idea, we wish to relate the disc correlators of
the two models on dynamical lattice.
More explicitly, we will find out the relation between
$D_L^\parallel$, $D_L^\perp$ of the $O(n)$ model and
$D_{a+L,\link{a\,a-1}}$, $D_{a-L,\link{a\,a+1}}$ of the RSOS model.
Our derivation of the relation is based on the loop equations
and the diagrammatics of the matrix models,
and does not rely on the continuum limit.

The first step is to relate the boundary cosmological constants.
From the relation (\ref{waw2}) we simply relate the $x$'s by
\begin{equation}
x_\text{RSOS}=x_\text{O(n)}
\end{equation}
for Neumann boundary of the $O(n)$ model and the fixed height
boundary in RSOS model.
Similarly, it follows from (\ref{yOn}) and (\ref{yRSOS}) that
the $y$'s for the $r$-th JS boundary and the $\link{ab}$ boundary
should be related via
\begin{align}
\begin{split}
y_\text{O(n)}&=\frac{y_\text{RSOS}}{S_a}
\quad\big(\link{a,b}=\link{r,r+1}~~\text{or}~~\link{h-r,h-r-1}\big),\\
y_\text{O(n)}&=\frac{y_\text{RSOS}}{S_b}
\quad\big(\link{a,b}=\link{r+1,r}~~\text{or}~~\link{h-r-1,h-r}\big).
\end{split}
\label{yy}
\end{align}

\subsection{Relations between Feynman graphs}

Instead of finding the correspondence of disc correlators quickly from
loop equations, let us explain how the correlators of the two
matrix models should be related from the viewpoint of Feynman
graph expansion.

The underlying idea is very simple.
Each Feynman graph of the RSOS matrix model describes a dynamical
lattice with a height assigned to every face, so that one can
draw contour lines separating the domains of adjacent heights.
If we focus only on the contour lines and forget about the heights,
then what we get is nothing but the Feynman graph of
the $O(n)$ matrix model.
We thus compare each Feynman graph of the $O(n)$ matrix model
with the sum over all the Feynman graphs of the RSOS model
having the same contour line configuration but different height assignments.

\subsubsection{Resolvents}

To begin with, let us consider the relation between the resolvents
of the RSOS and the $O(n)$ matrix models,
\begin{equation}
 W_a(x)= S_aW(x).
\label{waw3}
\end{equation}
Each graph contributing to $W_a(x)$ has the unique
outermost domain of height $a$, and is inscribed by
several subdomains of height $a\pm1$.
Each subdomain may be inscribed by several subdomains
of adjacent heights, and by iterating this a finite number of times
one can cover the entire disc.
Now let us sum over all the height assignments in the interior.
Using the rule explained after (\ref{flSOS}), we first assign $S_a$
to the outermost domain of height $a$.
To each of its subdomains one can assign the height $a+1$ or $a-1$,
which gives rise to a factor
\begin{equation}
  \frac{S_{a+1}}{S_{a}}+\frac{S_{a-1}}{S_{a}}=2\cos\frac\pi h =n.
\end{equation}
By repeating this and going step by step to the interior,
one ends up with the Feynman graph of the $O(n)$ matrix model
with a factor $n$ assigned to each loop.
This explains the relation (\ref{waw3}) at the level of the Feynman
graph sum.

\subsubsection{Disc correlators $D_0^\perp$ and $D_{a,\link{ab}}$}

Next, we use a similar argument to show the relation
\begin{equation}
D_{a,\link{ab}}(x,y)=D_0^\perp\big(x,y/S_a\big),
\label{dad0}
\end{equation}
where we assume $b=a+1$ for simplicity, and the JS boundary
is expected to be labelled by $r=a$ from (\ref{yy}).
To show this, we consider Feynman graphs contributing
to the LHS with the power $1/y^{m+1}$.
Such graphs have the $\link{ab}$-boundary of length $m$, and
there are therefore $m$ open contour lines ending on the
$\link{ab}$-boundary.
If we cut the graph along these contours, it would decompose into
$m_a$ pieces of boundary height $a$ and $m_b$ pieces
of boundary height $b$, where $m_a+m_b=m+1$ and $m_a>0$.
This means that the graphs have $m+1$ outermost domains of
heights $a$ or $b$.

We now take the sum of such graphs over
different height assignments in the interior
but for a fixed contour line configuration,
to obtain a graph of the $O(n)$ matrix model.
Each contour line in the interior is assigned a factor $n$
in the same way as before.
Collecting the factors associated to the outermost domains we find,
\begin{equation}
  y^{-m-1}S_a^{m_a}S_b^{m_b}~=~
  \big(S_a/y\big)^{m+1}
  \times \big(S_b/S_a\big)^{m_b}.
\label{mapD0}
\end{equation}

We then deform the contour line configuration to a loop configuration
by shrinking the $b$-part of the $\link{ab}$-boundary as shown
in Figure \ref{fig:rel1}.
\FIG{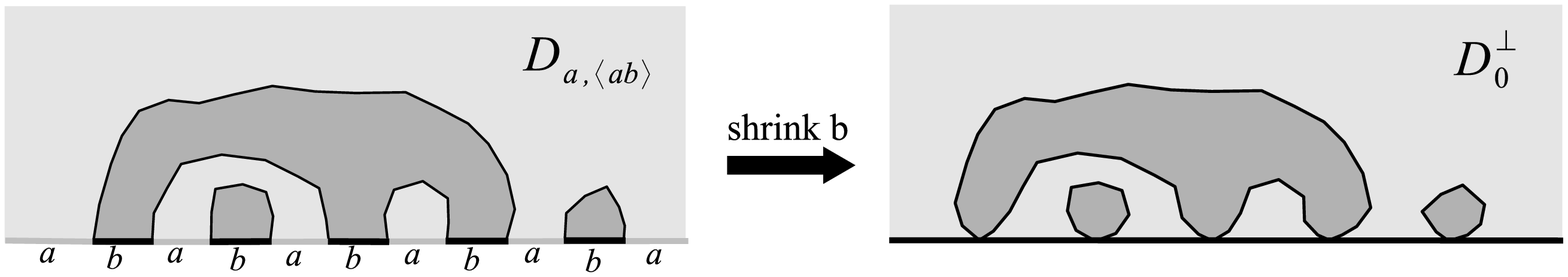}
    {The graphs contributing to $D_{a,\link{ab}}$ have
     open contour lines separating the domains of height $a$ and $b$.
     By shrinking the $b$ part of the $\link{ab}$-boundary,
     those contour lines turn to form loops touching the boundary.
     }{fig:rel1}
The $m$ open contour lines then turn into $m_b$ closed loops
touching the boundary $m$ times in total.
The expression for the weight (\ref{mapD0}) then implies that
the $\link{ab}$-boundary of the RSOS matrix model is mapped
to the $k$-th JS boundary of the $O(n)$ model, with boundary
cosmological constant $y/S_a$ and
\begin{equation}
k(a)=\frac{S_{a+1}}{S_a}=\frac{\sin{\pi(a+1)/h}}{\sin{\pi a/h}}.
\end{equation}
Thus we have shown (\ref{dad0}).
It also implies $a=r$ in agreement with (\ref{yy}).

\subsubsection{Disc correlators $D_1^\parallel$ and $D_{b,\link{ba}}$}

Using the same argument, let us next show the equation
\begin{equation}
 D_{b,\link{ba}}(x,y)=\frac{S_b}{y}
 \Big(W(x)+\frac1kD_1^{\parallel}\big(x,y/S_a\big)\Big).
\label{dbd1}
\end{equation}
We again focus on the graphs contributing to the LHS with
power $1/y^{m+1}$, which have $m$ open contour lines
ending on the $\link{ba}$-boundary.
Such graphs have $m_a$ outermost domains of height $a$
and $m_b$ outermost domains of height $b$, with the condition
\begin{equation}
 m_a+m_b=m+1,\qquad m_b>0.
\end{equation}
We perform the sum over the height assignments in the interior and
map the graphs to those of the $O(n)$ matrix model.

By shrinking the $b$-part of the $\link{ba}$-boundary, we get
the graph in the $O(n)$ matrix model which generically has
one open line connecting the two boundary changing operators
in addition to $m_b-1$ loops.
They altogether touch the JS boundary $m-1$ times in total.
It is important to notice that the open line can
end on the JS boundary.
We should therefore identify the boundary operators
as the one-leg blobbed operators $\mathbb{Y}_1^\parallel$.
The situation is illustrated in Figure \ref{fig:rel2}.
\FIG{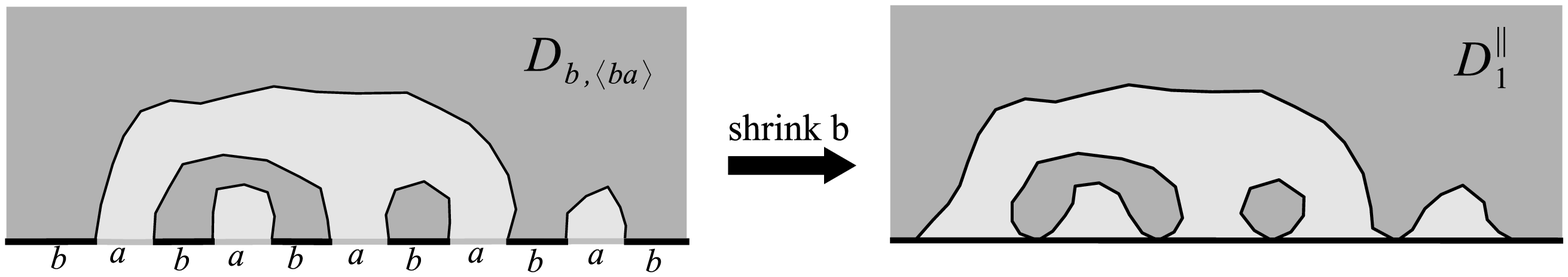}
    {The graphs contributing to $D_{b,\link{ba}}$ have open contour lines
     separating the domains of height $a$ and $b$.
     By shrinking the $b$ part of the $\link{ba}$-boundary, those contour lines
     turn into some loops touching the boundary and a line connecting
     the two boundary-changing operators.
     }{fig:rel2}
The graph of the $O(n)$ model we thus obtained has the
following weight from the outermost domains,
\begin{equation}
  y^{-m-1}S_a^{m_a}S_b^{m_b}~=~
  \big(S_b/y\big)\times
  \big(S_a/y\big)^m
  \times \big(S_b/S_a\big)^{m_b-1}.
\label{mapD1}
\end{equation}
The same graph and weight can be obtained from the Feynman graph
expansion of the second term in the RHS of (\ref{dbd1}).
Note that the additional factor $1/k$ is inserted because
the line connecting the two $\mathbb{Y}_1^\parallel$
can be made from propagators of $Y_1,\cdots$ or $Y_k$, leading to
a factor $k$.
The first term of the RHS, on the other hand,
is the sum over the exceptional graphs of the RSOS model
corresponding to $m=0$, namely, those graphs which have
the $\link{ba}$ boundary of length zero.
The sum over such graphs is simply the leading term in the
$1/y$-expansion of the LHS and therefore given by $W_b(x)/y$.
This finishes the diagrammatic proof of (\ref{dbd1}).

The relations (\ref{dad0}) and (\ref{dbd1}) can be used
to show that the loop equations of the $O(n)$ matrix model
(\ref{leOn0}) and the RSOS model (\ref{leRSOS}) are mapped to each other.
These relations also explain that the operators ${}^a\!B^{\link{a\,a+1}}$ and
${}^{a+1}\!B^{\link{a+1\,a}}$ have the same conformal weight as
that of $\mathbb{S}_0^\perp$ and $\mathbb{S}_1^\parallel$ between
the Neumann and the $k(a)$-th JS boundaries.

\subsubsection{Disc correlators of $L$-leg operators}

It is obvious how to extend the correspondence
to the disc correlators of $L$-leg operators using the argument
of summing over height configurations and shrinking the $b$-part
of the $\link{ab}$ boundary.
We skip the details and present the final results.
\begin{align}
\begin{split}
 D_{a-L,\link{a\,a+1}}(x,y) &=
 S_{a-1}\cdots S_{a-L}\;\frac{(n-k-L)!}{(n-k)!} D_L^\perp(x,y/S_a), \\
 D_{a+L,\link{a+1\,a}}(x,y) &=
 \frac{S_{a+1}\cdots S_{a+L}}{y}\Big(
 W_L(x)+ \frac{(k-L)!}{k!}D_L^\parallel(x,y/S_a)
 \Big),
\end{split}
\label{dldl}
\end{align}
where $W_L(x)$ is determined by the recursion relation
\begin{equation}
 \Disc W_{L+1}(x) = W_L(-x) \Disc W(x),\qquad
 W_1(x)=W(x).
\end{equation}
These include the results of the previous subsubsection as special cases.
It is also easy to show that they relate the recursion relation
of the $O(n)$ model (\ref{leOnL}) to that of the RSOS model
(\ref{recIntz}).
They also explain that the operator $^{a-L}\!B^{\link{a\,a+1}}$
has the same conformal weight as $\mathbb{S}_L^\perp$
between the Neumann and the $k(a)$-th JS boundaries,
and similarly for $^{a+L}\!B^{\link{a+1\,a}}$ and $\mathbb{S}_L^\parallel$.

The first term in the second line of (\ref{dldl}) is equal to
the leading term in the $1/y$-expansion of the LHS, and the
corresponding graphs have $\link{a+1\,a}$ boundary of zero length.
Therefore, $W_L(x)$ is a disc one-point function of a boundary
operator with $L$ nested loops attached, corresponding to
the fusion product of two $L$-leg operators.
Such an operator should be described in terms of
{\it star operators} \cite{Kostov:2003uh};
the star operator $S_L$ is a source of $L$ open lines and is
allowed to exist between two Neumann boundaries.
In \cite{Kostov:2003uh} its gravitational dimension was found to be
$\Delta_{1,L+1}$, and the disc two-point functions were computed
in the continuum limit.
Our $W_L$ can be calculated in the same way.
Using the standard parameterization $\xi=M\cosh\tau$
in the continuum limit we find,
\begin{equation}
 W_L(\tau)=\prod_{k=1}^L\frac{\sin\pi g}{\sin\pi g (k+1)}
           \prod_{k=0}^L W\big(\tau+i\pi(L-2k)\big).
\end{equation}

\section{Concluding remarks}\label{sec:concl}

In this paper we studied the new boundary condition
of the $O(n)$ model proposed by Jacobsen and Saleur.
By using the matrix model formulation, we were able to relate
them to the boundary conditions of RSOS model with alternating heights.
The loop equations turned out to be a very efficient tool
in calculating the spectrum and the conformal weights
of boundary changing operators.

Our techniques based on matrix model and loop equations are
applicable to the analysis of more involved situations,
such as discs with several JS boundaries labelled by different $k$.
An interesting problem is to study the spectrum of boundary operators
between two JS boundaries.
(On a regular lattice this has been done in the recent
 work \cite{dlh}.)

Another natural and interesting question will be to ask how our
results can be extended to the dilute phase.
Since the lattice will no longer be packed densely by the loops,
one would expect a conformal boundary condition for which
some sites on the boundary have no open line attached.
We will therefore need to generalize the JS boundary
so that it can have vacancies.
It will be an interesting problem to study the renormalization group
flow for the fugacity associated to the vacancy.
We hope to address this issue in the future.

\section*{Acknowledgments}

We would like to thank I. Kostov for helpful discussions and a
careful reading of the manuscript. One of the author (J.E.B.)
would like to thank J. Dubail for useful discussions.

\newpage
\appendix

\section{Gravitational dressing, Liouville theory and KPZ}\label{sec:kpz}

Here we summarize some basic facts on conformal field theories
coupled to gravity and boundary Liouville field theory.
More details can be found in \cite{Kostov:2003uh,Kostov:2002uq}
and the original paper
\cite{Fateev:2000ik, Ponsot:2001ng, Hosomichi:2001xc}.

Let us consider the `matter CFT' of the central charge
\begin{equation}
c=1-\frac{6\theta^2}{1-\theta} < 1.
\end{equation}
When $\theta=1/h$ for a positive integer $h\ge 3$, the theory is rational
and corresponds to the minimal model of the unitary series $(h-1,h)$.
Turning on the gravity corresponds to summing over different
metrics and topology of the two-dimensional space.
After gauge fixing it amounts to coupling the CFT to
the Liouville field $\phi$ and $bc$ ghost system.

Let us take the matter CFT to be the Coulomb gas model
\cite{Nienhuis:1982fx, Nienhuis:1984wm}
described by a scalar $\chi$.
The vanishing of the total central charge puts a certain condition
on the matter and the Liouville background charges $e_0$ and $Q$.
We can parameterize them as
\begin{equation}
Q   =\frac1\bL+\bL\,, \qquad
e_0 =\frac1\bL-\bL\,.
\end{equation}
Here $\bL =\sqrt{1-\theta}$ is the Liouville coupling constant.

In this model we consider the matter field $e^{ie_{r,s}\chi}$
of conformal weight $\delta_{r,s}$.
With a suitable gravitational dressing it becomes an operator of conformal
weight one,
\begin{equation}
 B_{r,s}=\frac{\Gamma(2\bL P_{r,s})}{\pi}
 \exp\left(ie_{r,s}\chi+\beta_{r,s}\phi\right).
\end{equation}
Here various parameters are related as follows,
\begin{equation}
 e_{r,s}=\frac{e_0}2-P_{r,s},\qquad
 \beta_{r,s}=\frac Q2-|P_{r,s}|,\qquad
 P_{r,s}=\frac{r}{2\bL}-\frac{s\bL}2,
\end{equation}
and $(r,s)$ is a pair of positive integers labelling degenerate
representations of the matter CFT.
The matter conformal weight $\delta_{r,s}$ is given by
\begin{equation}
\delta_{r,s}=\frac{(r/\bL -s\bL )^2-(1/\bL -\bL )^2}{4}\,.
\end{equation}
Introducing {\it gravitational dimension} $\Delta\equiv (2P-e_0)/2\bL$
and {\it string susceptibility} $\gamma_\text{str}\equiv-\theta/(1-\theta)$,
one can write the KPZ scaling formula
\cite{Knizhnik:1988ak,David:1988hj,Distler:1988jt},
\begin{equation}
\delta=\frac{\Delta(\Delta-\gamma_\text{str})}{1-\gamma_\text{str}}.
\label{EquKPZ}
\end{equation}
Note that there is another way of gravitational dressing,
$\tilde\beta=Q/2+|P|$, as was considered in \cite{Kostov:2007jj}.
As an example, the boundary identity operator can be dressed by
$e^{\phi/\bL}$ instead of $e^{\bL\phi}$.
This suggests that there are two boundary cosmological couplings,
and one has a fractional dimension with respect to the other.

After turning on the gravity, correlators no longer depend on the
positions of the operators inserted because one has to integrate
over the positions of those operators.
The dimensions of the operators therefore cannot be read from
the position-dependence of their correlators.
Instead, they should be read off from the dependence
of correlators on the cosmological constant $\mu$.
If we restrict to discs, then the amplitudes with $n$ boundary
operators $B_{P_i}$ and $m$ bulk operators $V_{K_j}$ scale with $\mu$ as
\begin{equation}
 \big\langle {}^{\xi_1}\!B_{P_1}\!\!{}^{\xi_2}\cdots
            {}^{\xi_n}\!B_{P_n}\!\!{}^{\xi_1}\;
  V_{K_1}\cdots V_{K_m} \big\rangle \propto \mu^\gamma,
\end{equation}
with
\begin{equation}
 \gamma=\big(1-m-\frac n2\big)
       \big(1-\frac{\gamma_\text{str}}{2}\big)
 +\frac{1}{2\bL }\Big(\sum_{i=1}^n{|P_i|}+\sum_{j=1}^m{|K_j|}\Big).
\label{muscal}
\end{equation}
Of course, the gravitational dimensions of the operators can be
read more explicitly from the more detailed form of the amplitudes.

As an example, let us consider a disc with two boundary segments,
labelled by boundary cosmological constants $\zeta_1$ and $\zeta_2$
and connected by the operators ${}^{\zeta_1}\!B^{\zeta_2}$.
In boundary Liouville theory, boundary cosmological constant $\zeta$
is the coefficient of the boundary interaction $e^{\bL\phi}$.
Following \cite{Fateev:2000ik} we use a parametrization of $\zeta$ similar to
(\ref{zetas}),
\begin{equation}
\zeta=\sqrt{\frac{\mu}{\sin{\pi\bL^2}}}\cosh{\bL^2\tau}.
\end{equation}
The computation of the disc amplitude involves the disc two-point
functions of the Liouville and matter CFTs.
The Liouville and matter part of the correlator factorize,
and the matter part gives only a $\zeta$-independent factor.
The Liouville part is given by \cite{Fateev:2000ik}
\begin{equation}
\big\langle {}^{\zeta_1}\!B_P{}^{\zeta_2}(x)\,
            {}^{\zeta_2}\!B_P{}^{\zeta_1}(0)\big\rangle_\text{Liouville}
 = A(P)d(|P|,\tau_1,\tau_2)|x|^{-2\beta(Q-\beta)}\,,
\label{2points}
\end{equation}
with
\begin{align}
\begin{split}
\ln d(P,\tau_1,\tau_2)=
 -\int_{-\infty}^\infty
 \frac{d\omega}{\omega}\left(
 \frac{\cos\omega\tau_1\cos\omega\tau_2\sinh 2\pi P\omega/\bL}
      {\sinh\pi\omega\sinh\pi\omega/\bL^2}
 -\frac{2P\bL}{\pi\omega}\right).
\end{split}
\label{exprd}
\end{align}
Here $A(P)$ is a known function of $P$ and is related to
the ``leg factor'' arising from different normalization of the
wave functions.
It is independent of $\tau$'s and therefore unimportant.
On the other hand, the $\tau$-dependent part (\ref{exprd}) is
expressed in terms of the double-sine function
of pseudo-periods $\bL$ and $1/\bL$ \cite{Kharchev:2001rs}.
It satisfies an important shift relation involving
both $P$ and $\tau_i$,
\begin{equation}
 d(P,\tau_1+i\pi,\tau_2)-d(P,\tau_1-i\pi,\tau_2)
 \;\propto\; \sinh{\bL^2\tau_1}\, d(P-\bL/2,\tau_1,\tau_2),
\label{shiftFZZ}
\end{equation}
up to a $\tau$-independent factor.
Shifting $P$ by $-\bL/2$ corresponds to changing the label
of the operator from $(r,s)$ to $(r,s+1)$.

\section{Solving the loop equation}\label{sec:sol}

Here we solve the loop equation (\ref{d0d1le})
\begin{align*}
   d_1\big(\tau\mp\frac{i\pi}2,\sigma\big)
   d_0\big(\tau\pm\frac{i\pi}2,\sigma\big)
  = CM^g\cosh\frac{g(\tau+\sigma)\pm\alpha}{2}
        \cosh\frac{g(\tau-\sigma)\pm\alpha}{2}\;.
\end{align*}
We define the function $u_a(\tau,\sigma)$ by
\begin{equation}
d_a(\tau,\sigma)=\frac{\sqrt{C}}2M^{\alpha_a}\exp u_a(\tau,\sigma)
\end{equation}
with $\alpha_a+\alpha_b=g$, and denote by $\hat u_a(\omega,\sigma)$
their Fourier transform with respect to $\tau$.

Let us take the log and the Fourier transform of the loop equation.
Using
\begin{equation}
 \int_{-\infty}^\infty d\tau e^{i\omega\tau}
 \log\Big(2\cosh\frac{g\tau+\alpha}2\Big)
 = -\frac{\pi e^{-i\alpha\omega/g}}{\omega\sinh(\pi\omega/g)},
\end{equation}
the loop equation for $\hat u_a(\omega,\sigma)$ becomes algebraic,
\begin{equation}
  e^{\pm\frac{\pi\omega}2}\hat u_1(\omega,\sigma)
 +e^{\mp\frac{\pi\omega}2}\hat u_0(\omega,\sigma)
 \;=\; -\frac{2\pi}\omega\frac{\cos\omega\sigma}{\sinh(\pi\omega/g)}
 e^{\pm i\omega\alpha/g}.
\end{equation}
Solving this in favor of $\hat u_a$ and Fourier transforming back,
we find
\begin{equation}
 u_a(\tau,\sigma) ~=~
-\int_{-\infty}^\infty \frac{d\omega}{\omega}
 \frac{\cos\sigma\omega\cos\tau\omega}
      {\sinh\pi\omega\sinh\pi\omega/g}
 \sinh\Big(\frac{\pi\omega}2\pm\frac{i\alpha\omega}g\Big),
\label{ua}
\end{equation}
where plus sign is for $u_1$ and minus sign for $u_0$.
One recognizes the same functional form as the Liouville
boundary two-point function (\ref{exprd}).

By comparing (\ref{ua}) with (\ref{exprd}) one finds the value of $P$
for the boundary-changing operators.
Then by using the scaling law (\ref{muscal}) one can determine
the exponents $\alpha_0$ and $\alpha_1$
\begin{equation}
 2\bL P_0 = \alpha_0 = r\theta, \qquad
 2\bL P_1 = \alpha_1 = 1-\theta-r\theta.
\label{aa}
\end{equation}
Another way to find $\alpha_a$ is to analyze the two-point
function at $\zeta=\mu=0$,
\begin{equation}
d(\xi,\zeta=0,\mu=0)\sim\xi^{\alpha_a}\sim e^{\alpha_a|\tau|}.
\label{pe3}
\end{equation}
Setting $\sigma=i\pi/2g$ and $\tau\to\infty$ in (\ref{ua}),
the dominant contribution to the integral is from the vicinity
of the second order pole at $\omega=0$.
Using
\begin{equation}
\int_{-\infty}^\infty \frac{d\omega e^{i\omega\tau}}{\omega^2}
 = -\pi|\tau|,
\end{equation}
one finds $u_a \simeq 2\bL P|\tau|$ and recovers (\ref{aa}) again.

\end{document}